\documentclass[11pt,german]{article}

\columnwidth\textwidth

\def\semicolon{\nobreak\mskip2mu\mathpunct{}\nonscript\mkern-\thinmuskip{;}
\mskip6muplus1mu\relax} 

\def\Id{
{\rm Id}\, }

\def\det{
{\rm det}}

\newcommand{\e}{\mathrm{e}}
\renewcommand{\i}{\mathrm{i}}

\usepackage{amsmath}
\usepackage{amssymb}
\usepackage{times}
\usepackage{dsfont}

\begin{document}

\thispagestyle{plain}

\title{Complex representation theory of the
electromagnetic field}

\author{Andreas Aste\\
Paul Scherrer Institute, CH-5232 Villigen PSI, Switzerland}
\date{November 2, 2012}

\date{November 30, 2012}

\maketitle

\begin{abstract}
A concise discussion of the 3-dimensional irreducible (1,0) and (0,1) representations
of the restricted Lorentz group and their application to the description of the
electromagnetic field is given. It is shown that a mass term is in conflict with
relativistic invariance of a formalism using electric and magnetic fields only,
contrasting the case of the two-component Majorana field equations.
An important difference between the Dirac equation and the Dirac form of Maxwell's equations
is highlighted by considering the coupling of the electromagnetic field to the electric current.
\end{abstract}

\label{first}

\section{Introduction}
Starting from Lorentz symmetry as the key property of Minkowski
space-time in the framework of the special theory of relativity,
we may observe that the classical electric and magnetic field
can be combined into a single "\emph{photon wave function}" \cite{Birula}
\begin{equation}
\Psi=\frac{1}{\sqrt{2}} (E+\i B) \, , \quad \i^2=-1  \label{normdef}
\end{equation}
where the electric field E and the magnetic field B are three-component
real fields which, for the sake of convenience, shall be written in column matrix form
\begin{equation}
E(x)=
\left(\begin{array}{r}
E_1(x) \\
E_2(x) \\
E_3(x)
\end{array}\right)  ,  \; \,
B(x)=
\left(\begin{array}{r}
B_1(x) \\
B_2(x) \\
B_3(x) 
\end{array}\right)  ,  \; \, 
x = \{x^\mu\}=
\left(\begin{array}{r}
x^0 \\
x^1 \\
x^2 \\
x^3
\end{array}\right) 
\end{equation}
in the following. The column vector $x$ denotes Cartesian space-time coordinates
$ x= (x^0=ct,x^1,x^2,x^3)^{\mbox{\tiny{T}}}=$
$(x^0,{\bf{x}}^{\mbox{\tiny{T}}})^{\mbox{\tiny{T}}}=$
$(x_0,-x_1,-x_2,-x_3)^{\mbox{\tiny{T}}}$ in an orthonormal standard coordinate system in Minkowski space.
Throughout the paper, we will choose a system of units where the speed of light is $c=1$.\\

\noindent Hence, the Maxwell-Faraday equation
\begin{equation}
\frac{\partial B}{\partial t}= - \nabla \times E = - curl \, E 
\end{equation}
and Amp\`ere's circuital law \emph{in vacuo}
\begin{equation}
\frac{\partial E}{\partial t} = +\nabla \times B \label{ampere}
\end{equation}
can be cast into one single Lorentz-covariant equation of motion
\begin{equation}
\frac{\partial \Psi}{\partial t}  = -\i \cdot  \nabla \times \Psi \, . \label{fun}
\end{equation}
This was already recognized in lectures by Riemann in the nineteenth century \cite{Weber}.
A short related note can also be found in the lecture notes of Sommerfeld \cite{Sommerfeld}.

\section{Field Equations}

\noindent Taking the divergence of equation (\ref{fun}) 
\begin{equation}
\nabla \cdot \dot{\Psi} = -\i \cdot \nabla \cdot (\nabla \times \Psi) = 0
\end{equation}
readily shows that the divergence of the electric and magnetic field is conserved.
Therefore, if the analytic condition $ div \, E = div \,  B=0$ holds due to the absence of electric or magnetic charges
on a space-like slice of space-time, it holds everywhere.\\

\noindent The field equations single out the helicity eigenstates of the photon wave function
which are admissible for massless particles
according to Wigner's analysis of the unitary representations of the Poincar\'e group \cite{Wigner}.
E.g., a circularly polarized (right-handed) plane wave moving in {\emph{positive}} $x^3$-direction is given by
\begin{equation}
\Psi_R(x)=N(k^0) \left(\begin{array}{r}
1 \\
\i \\
0 \\
\end{array}\right) e^{\i k^3 x^3 -\i k^0 x^0}
=N(k^0) (\hat{e}_1+ \i \hat{e}_2)e^{\i k^3 x^3 - \i k^0 x^0} \, \, , \quad k^0=k^3 > 0
\end{equation}
where $N(k^0)$ is a normalization factor, whereas the corresponding left-handed plane wave is given by
\begin{equation}
\Psi_L(x)=N(k^0) \left(\begin{array}{r}
1 \\
\i \\
0 \\
\end{array}\right) e^{- \i k^3 x^3 + \i k^0 x^0} \, \, , \quad k^0=k^3 > 0 \, .
\end{equation}
If the right-handed wave moves in negative $x^3$-direction ($k^3<0$), one has
\begin{equation}
\Psi_R(x)=N(k^0) \left(\begin{array}{r}
1 \\
-\i \\
0 \\
\end{array}\right) e^{\i k^3 x^3 - \i k^0 x^0} \, \, , \quad k^0=|k^3| > 0 \, .
\end{equation}

\noindent The presence of electric charges and the absence of magnetic charges
breaks the gauge symmetry of equation (\ref{fun})
\begin{equation}
\Psi \mapsto \e^{\i \alpha} \Psi \, , \quad \alpha \in \mathbb{R} \, .
\end{equation}

\noindent Introducing antisymmetric matrices $\tilde \Sigma_1$, $\tilde \Sigma_2$, and
$\tilde \Sigma_3$ defined by the help
of the totally antisymmetric tensor in three dimensions $\varepsilon$ fulfilling
$\varepsilon_{123}=1$, $\varepsilon_{lmn}=-\varepsilon_{mln}
=-\varepsilon_{lnm}$ as follows
\begin{equation}
(\tilde \Sigma_l)_{mn} = \i \varepsilon_{lmn}
\end{equation}
\begin{equation}
\tilde \Sigma_1=\left(\begin{array}{cccc}
0 & 0 & 0 \\
0 & 0 & \i \\
0 & -\i & 0 
\end{array}\right)\; , \quad
\tilde \Sigma_2=\left(\begin{array}{cccc}
0 & 0 & -\i \\
0 & 0 & 0 \\
\i & 0 & 0 
\end{array}\right)\; , \quad
\tilde \Sigma_3=\left(\begin{array}{cccc}
0 & \i & 0 \\
-\i & 0 & 0 \\
0 & 0 & 0 
\end{array}\right)\; 
\end{equation}
equation (\ref{fun}) can be written in the form ($\partial_k = \partial / \partial x^k \, ,$ $k=1,2,3$)
\begin{equation}
\frac{\partial \Psi}{\partial t} = \tilde \Sigma_k \partial_k \Psi 
\end{equation}
or, defining matrices $\Gamma^\mu$ by $\Gamma^0={\Id}_3$, where ${\Id}_3$ denotes the 
$3 \times 3$  identity matrix, and $\Gamma_k= \tilde \Sigma_k=-\Gamma^k$ for $k=1,2,3,$ equation (\ref{fun})
finally reads
\begin{equation}
\i \Gamma^\mu \partial_\mu \Psi = 0 \, . \label{nice}
\end{equation}
Obviously, the complex conjugate wave functions $\Psi^*$ fulfills the equation
\begin{equation}
\i \bar \Gamma^\mu \partial_\mu \Psi^* = 0 
\end{equation}
where $\bar \Gamma^\mu = ({\Id}_3, \tilde \Sigma_1,
\tilde \Sigma_2, \tilde \Sigma_3)$.\\

\noindent Defining by the help of the hermitian conjugate field $\Psi^+ = \Psi^{* \mbox{\tiny{T}}}$
the four density components
\begin{equation}
T^{0 \mu} = \Psi^+ \Gamma^\mu \Psi
\end{equation}
we recover after a short calculation the energy density and the Poynting vector of the
electromagnetic field
\begin{equation}
\omega=T^{00}=\frac{1}{2} ( E^2 + B^2 ) \, , \quad s^k = T^{0k}= (E \times B)_k \, . \label{density}
\end{equation}\\
Equations (\ref{density}) signal a crucial difference between the Dirac equation for a
spin-$\frac{1}{2}$ (anti-) particle with mass m
\begin{equation}
\i \gamma^{\mu} \partial_\mu \psi - m \psi = 0
\end{equation}
and equation (\ref{nice}), since the (probability) four-current density
\begin{equation}
j^\mu_{Dirac} = \psi^+ \gamma^0 \gamma^\mu \psi = \bar{\psi} \gamma^\mu \psi
\end{equation}
transforms a a vector field, whereas $T^{0 \mu}=T^{\mu 0}$ is related to the electromagnetic
stress-energy tensor $T^{\mu \nu}$.

\noindent $\Psi$ and $\Psi^*$ transform according to the (1,0) and (0,1) = (1,0)$^*$ representation
of the proper Lorentz group (see below).
We mention as a historical fact that Fredrik Jozef Belinfante coined the expression \emph{undor}
when dealing with fields transforming according to some specific representations of the Lorentz group.
Here, we term the complex Riemann-Silberstein three-component field
$\Psi$ a \emph{bivector field} \cite{Silberstein},
in order to allow for a clear distinction from vector or spinor fields. Furthermore, this term was
already used by Ludwig Silberstein \cite{Silberstein} in 1907.

\section{Transformation Properties}
\noindent E and B transform as vectors under spatial rotations
$ (x^0, {\bf{x}}') \rightarrow (x'^0, {\bf{x}}')=(x^0, R \bf{x})$
according to
\begin{equation}
C'(x')=R C(x)
\end{equation}
where R is an orientation preserving rotation matrix in the
special orthogonal group
\begin{equation}
{\rm{SO}}(3)=\{R \! \in \! {\rm{Mat}}(3,\mathbb{R}) \, \semicolon 
R^{\mbox{\tiny{T}}}  R=+\Id_{3},  det \, R = 1\} \, .
\end{equation}

\noindent The definition of ${\rm{SO}}(3)$ is rooted in the
preservation of the Euclidean scalar pro\-duct of real three-vectors $({\bf{x}},{\bf{y}})={\bf{x}}^{\mbox{\tiny{T}}} {\bf{y}} =
 (R{\bf{x}},R{\bf{y}}) =
{\bf{x}}^{\mbox{\tiny{T}}} R^{\mbox{\tiny{T}}} R {\bf{y}}$. From $R^{\mbox{\tiny{T}}} R=
{\Id}_3$ follows $det \, R = \pm 1$; the additional condition $det \, R=+1$
excludes spatial reflections from ${\rm{SO}}(3)$.\\

\noindent However, the electromagnetic field components are not related to spatial components of
a four-vector with respect to the proper Lorentz group
\begin{equation}
{\rm{SO}}^+(1,3)=\{ \Lambda \! \in \! {\rm{Mat}}(4,\mathbb{R}) \, \semicolon 
\Lambda^{\mbox{\tiny{T}}} g \Lambda=g, \,
\Lambda^0_{\, \, 0} \ge 1, \, \det \,  \Lambda = +1\} 
\end{equation}
with the metric tensor g defined according to the sign convention given by
\begin{equation}
g=
\mbox{diag}(1,-1,-1,-1)=
\left(\begin{array}{rrrr}
 1 &  0 & 0  & 0 \\
 0 & -1 & 0  & 0 \\
 0 &  0 & -1 & 0 \\
 0 &  0 & 0  & -1 
\end{array}\right) \; .
\end{equation}
The explicit representation of $\Lambda \! \in \! {\rm{SO}}^+(1,3)$ by its matrix elements is given by
\begin{equation}
\Lambda=
\left(\begin{array}{rrrr}
 \Lambda^0_{\, \, 0} &   \Lambda^0_{\, \, 1} &  \Lambda^0_{\, \, 2}  &  \Lambda^0_{\, \, 3} \\
 \Lambda^1_{\, \, 0} &   \Lambda^1_{\, \, 1} &  \Lambda^1_{\, \, 2}  &  \Lambda^1_{\, \, 3} \\
 \Lambda^2_{\, \, 0} &   \Lambda^2_{\, \, 1} &  \Lambda^2_{\, \, 2}  &  \Lambda^2_{\, \, 3} \\
 \Lambda^3_{\, \, 0} &   \Lambda^3_{\, \, 1} &  \Lambda^3_{\, \, 2}  &  \Lambda^3_{\, \, 3} 
\end{array}\right) \; .
\end{equation}

\noindent Still, the electromagnetic field $\Psi$ transforms as a vector under the complex
special orthogonal group in three dimensions
\begin{equation}
{\rm{SO}}(3,\mathbb{C})=\{Q \! \in \! {\rm{Mat}}(3,\mathbb{C}) \, \semicolon  
Q^{\mbox{\tiny{T}}}  Q=+\Id_{3},  det \, Q = 1\} \, .
\end{equation}
This observation is related to the fact that the proper Lorentz group and the complex rotation group
${\rm{SO}}(3,\mathbb{C})$ are isomorphic indeed
\begin{equation}
{\rm{SO}}^+(1,3) \cong {\rm{SO}}(3,\mathbb{C}) \, .
\end{equation}
\noindent The elegance and the analogy of the considerations presented so far
to the Dirac \cite{Dirac}, Weyl \cite{Weyl} or \emph{massive} two-component
Majorana formalism \cite{Majorana} is obvious.
Still, a mass term is absent in equation (\ref{nice}). It is one purpose of this paper to explicitly
show that such a term cannot be established, which enforces a new concept like gauge theories when
massive spin 1 particles are involved in a field theory.

\section{Symmetries: Generators of ${\rm{SO}}^+(1,3)$, ${\rm{SO}}(3)$ and ${\rm{SO}}(3,\mathbb{C})$}
A pure Lorentz boost in $x^1$-direction with velocity
$\beta=\beta_1 $ and Lorentz factor $\gamma=\gamma_1$ is expressed by the
matrix
\begin{equation}
\Lambda =\left(\begin{array}{cccc}
\gamma & -\gamma \beta & 0 & 0 \\
-\gamma \beta & \gamma & 0 & 0 \\
0 & 0 & 1 & 0 \\
0 & 0 & 0 & 1 
\end{array}\right) \;, \quad \gamma=\frac{1}{\sqrt{1-\beta^2}}, \quad
\gamma^2-\gamma^2 \beta^2=1.
\end{equation}
which can be written to first order in $\beta$ as
\begin{equation}
\Lambda =\left(\begin{array}{cccc}
1 & 0 & 0 & 0 \\
0 & 1 & 0 & 0 \\
0 & 0 & 1 & 0 \\
0 & 0 & 0 & 1 
\end{array}\right)+
\beta
\left(\begin{array}{cccc}
0 & -1 & 0 & 0 \\
-1 & 0 & 0 & 0 \\
0 & 0 & 0 & 0 \\
0 & 0 & 0 & 0 
\end{array}\right) =1+\beta L_1 
\end{equation}
where $L_1$ is a generator for boosts in $x^1$-direction.
The original Lorentz boost is recovered by
exponentiating the generator multiplied by the boost parameter $\xi_1$
\begin{equation}
\exp (\xi_1 L_1)=\left(\begin{array}{cccc}
+\cosh \xi_1 & -\sinh \xi_1 & 0 & 0 \\
-\sinh \xi_1 & +\cosh \xi_1 & 0 & 0 \\
0 & 0 & 1 & 0 \\
0 & 0 & 0 & 1 
\end{array}\right) \; , \quad \cosh(\xi_1)=\gamma_1 \, . \label{expo1}
\end{equation}
Additional generators $L_2$ and $L_3$ for boost in $x^2$- and $x^3$-direction are
\begin{equation}
L_2=\left(\begin{array}{cccc}
0 & 0 & -1 & 0 \\
0 & 0 & 0 & 0 \\
-1 & 0 & 0 & 0 \\
0 & 0 & 0 & 0 
\end{array}\right)\; , \quad
L_3=\left(\begin{array}{cccc}
0 & 0 & 0 & -1 \\
0 & 0 & 0 & 0 \\
0 & 0 & 0 & 0 \\
-1 & 0 & 0 & 0 
\end{array}\right)
\end{equation}
and generators for rotations around the $x^1$-, $x^2$-, and
$x^3$-axis are
\begin{equation}
S_1=\left(\begin{array}{cccc}
0 & 0 & 0 & 0 \\
0 & 0 & 0 & 0 \\
0 & 0 & 0 & +1 \\
0 & 0 & -1 & 0 
\end{array}\right) , \;
S_2=\left(\begin{array}{cccc}
0 & 0 & 0 & 0 \\
0 & 0 & 0 & -1 \\
0 & 0 & 0 & 0 \\
0 & +1 & 0 & 0 
\end{array}\right) , \;
S_3=\left(\begin{array}{cccc}
0 & 0 & 0 & 0 \\
0 & 0 & +1 & 0 \\
0 & -1 & 0 & 0 \\
0 & 0 & 0 & 0 
\end{array}\right)\, . 
\end{equation}
Altogether, these six generators of the proper Lorentz group span the Lie algebra
$\frak{so}^+(1,3)$, satisfying
the commutation relations
\begin{equation}
[S_l,S_m]=-\varepsilon_{lmn} S_n, \quad
[L_l,L_m]=+\varepsilon_{lmn} S_n, \quad
[L_l,S_m]=-\varepsilon_{lmn} L_n
\end{equation}
with the totally antisymmetric tensor or ${\rm{SO}}(3)$-structure constants $\varepsilon$
in three dimensions. Note that generators are often
multiplied with the imaginary unit $\i$ in the physics literature
in order to get Hermitian matrices.

\noindent Restricting our considerations to the rotation group ${\rm{SO}}(3)$ only, a basis of the
Lie Algebra $\frak{so}(3)$ is given by $(\Sigma_l)_{mn}=\varepsilon_{lmn}$, or explicitly
\begin{equation}
\Sigma_1=\left(\begin{array}{cccc}
0 & 0 & 0 \\
0 & 0 & +1 \\
0 & -1 & 0 
\end{array}\right)\; , \quad
\Sigma_2=\left(\begin{array}{cccc}
0 & 0 & -1 \\
0 & 0 & 0 \\
+1 & 0 & 0 
\end{array}\right)\; , \quad
\Sigma_3=\left(\begin{array}{cccc}
0 & +1 & 0 \\
-1 & 0 & 0 \\
0 & 0 & 0 
\end{array}\right)
\end{equation}
with
\begin{equation}
[\Sigma_l,\Sigma_m]=-\varepsilon_{lmn} \Sigma_n \, . 
\end{equation}
By decomposing a real rotation matrix according to $R = {\Id}_3 + \delta R$, we obtain in a straightforward
manner
\begin{equation}
R^{\mbox{\tiny{T}}} R = ({\Id}_3 + \delta R)^{\mbox{\tiny{T}}} ({\Id}_3 + \delta R)= {\Id}_3 +\delta R
^{\mbox{\tiny{T}}} + \delta R + \delta R^{\mbox{\tiny{T}}}  \delta R = {\Id}_3 \, . \label{gen}
\end{equation}
For \emph{small} $\delta R$, $\delta R^{\mbox{\tiny{T}}}  \delta R$ is negligible and
$\delta R^{\mbox{\tiny{T}}}+ \delta R =0$ holds approximately,
correspondingly the generators in $\frak{so}(3)$ must be anti\-sym\-metric. Therefore, the real and
antisymmetric $\Sigma-$matrices form
a basis of $\frak{so}(3)$.

\noindent By definition, the same argument given by equation (\ref{gen}) holds for the complex group ${\rm{SO}}(3,\mathbb{C})$.
A complete basis of the Lie algebra $\frak{so}(3,\mathbb{C})$ is thus obtained by adding the antisymmetric
matrices $\tilde \Sigma_k = \i \Sigma_k$ to the generators $\Sigma_1$, $\Sigma_2$, and $\Sigma_3$
of the real rotation group ${\rm{SO}}(3)$. This complexification leads to
\begin{equation}
[\Sigma_l,\Sigma_m]=-\varepsilon_{lmn} \Sigma_n \, , \quad 
[\tilde \Sigma_l, \tilde \Sigma_m]=+\varepsilon_{lmn} \Sigma_n \, , \quad
[\tilde \Sigma_l,\Sigma_m]=-\varepsilon_{lmn} \tilde \Sigma_n
\end{equation}
i.e. the same abstract Lie algebra is obtained if one identifies the generators  of ${\rm{SO}}^+(1,3)$
and ${\rm{SO}}(3,\mathbb{C})$ according to $S_l \leftrightarrow \Sigma_l$ and 
$L_l \leftrightarrow \tilde \Sigma_l$ for $ l=1,2,3$.

\noindent An arbitrary matrix $\Lambda \in$ ${\rm{SO}}^+(1,3)$ can be written in the form
\begin{equation}
\Lambda=\exp(\xi_1 L_1+\xi_2 L_2 + \xi_3 L_3 + \alpha_1 S_1 + \alpha_2 S_2 + \alpha_3 S_3) 
\end{equation}
establishing a one-one correspondence to $Q \in {\rm{SO}}(3,\mathbb{C})$ via
\begin{equation}
Q =\exp(\xi_1 \tilde \Sigma_1+\xi_2 \tilde \Sigma_2 + \xi_3 \tilde \Sigma_3 +
\alpha_1 \Sigma_1 + \alpha_2 \Sigma_2 + \alpha_3 \Sigma_3) \, .
\end{equation}
E.g., one has in correspondence to equation (\ref{expo1}) for a bivector boost in x$^1$-direction
an ${\rm{SO}}(3,\mathbb{C})$ transformation matrix
\begin{equation}
\exp(\xi_1 \tilde \Sigma_1) = 
\left(\begin{array}{cccc}
1 & 0 & 0 \\
0 & \cosh \xi_1 & +\i \sinh \xi_1 \\
0 & -\i \sinh \xi_1 & \cosh \xi_1 
\end{array}\right)\; =
\left(\begin{array}{cccc}
1 & 0 & 0 \\
0 & \gamma_1 & +\i \gamma_1 \beta_1 \\
0 &  -\i \gamma_1 \beta_1 & \gamma_1
\end{array}\right)\;
\end{equation}
and a rotation around the x$^1$-axis is obtained by acting on the bivector with
\begin{equation}
\exp(\alpha_1  \Sigma_1) = 
\left(\begin{array}{cccc}
1 & 0 & 0 \\
0 & \cos \alpha_1 & + \sin \alpha_1 \\
0 & - \sin \alpha_1 & \cos \alpha_1 
\end{array}\right)\; \in {\rm{SO}}(3) \, .
\end{equation}
A short exercise shows that acting with $\exp(\xi_1 \tilde \Sigma_1)$ on the bivector field
generates the correct Lorentz transformations of the electromagnetic field derived
in many standard textbooks for a boost in x$^1$-direction
\begin{equation}
\left(\begin{array}{c}
E'_1+ \i B'_1 \\
E'_2+ \i B'_2 \\
E'_3+ \i B'_3 
\end{array}\right)\; =
\left(\begin{array}{cccc}
1 & 0 & 0 \\
0 & \gamma_1 & +\i \gamma_1 \beta_1 \\
0 &  -\i \gamma_1 \beta_1 & \gamma_1
\end{array}\right)\;
\left(\begin{array}{c}
E_1+ \i B_1 \\
E_2+ \i B_2 \\
E_3+ \i B_3 
\end{array}\right)\;
\end{equation}
therefore
\begin{displaymath}
E'_1=E_1 \, , \quad B'_1=B_1 
\end{displaymath}
\begin{displaymath}
E'_2=\gamma_1 (E_2 - \beta_1 B_3) \, , \quad B'_2=\gamma_1 (B_2+ \beta_1 E_3) 
\end{displaymath}
\begin{equation}
E'_3=\gamma_1 (E_3 + \beta_1 B_2) \, , \quad B'_3=\gamma_1 (B_3- \beta_1 E_2) \, .
\end{equation}

\noindent Generally, one has in the case of a Lorentz transformation
$x'^\mu = \Lambda^\mu_{\, \, \nu} x^\nu$
\begin{equation}
\Psi'(x')=Q \Psi(x) = Q \Psi(\Lambda^{-1} x') \, .
\end{equation}
Note that the transformation property of $\Psi$ implies
that the real and imaginary part of
\begin{equation}
\Psi^{\mbox{\tiny{T}}} \Psi = \Psi^{\mbox{\tiny{T}}} Q^{\mbox{\tiny{T}}} Q \Psi =
 \Psi^{\mbox{\tiny{T}}} Q^{-1} Q \Psi = \frac{1}{2}(E^2-B^2) + \i E \cdot B
\end{equation}
are Lorentz-invariant quantities.

\section{Mass Terms}
\subsection{Majorana Approach}
Acting with the operator $-\i\bar \Gamma^\mu \partial_\mu$ on the field equation (\ref{nice}) leads to
\begin{equation}
\bar \Gamma_\mu \Gamma_\nu \partial_\nu \partial_\mu \Psi = \partial_0^2+\nabla \times \nabla \times \Psi=
 \partial_0^2 \Psi -\Delta \Psi +\nabla  (\nabla \cdot \Psi)=0 
\end{equation}
therefore the wave function $\Psi$ fulfills the free wave equation
\begin{equation}
\Box \Psi = \partial^\mu \partial_\mu \Psi=0
\end{equation}
in the absence of charges, ensuring the correct relativistic energy-momentum relation.
Introducing a naive mass term for the $\Psi$-field like
\begin{equation}
\i \Gamma^\mu \partial_\mu \Psi - m \Psi=0
\end{equation}
would spoil the relativistic invariance of the field equation.
As a more general approach one may introduce an (anti-)linear operator S and make the ansatz
\begin{equation}
\i \Gamma^\mu \partial_\mu \Psi - m S \Psi=0  \, .
 \end{equation}
Since plane-wave solutions $\sim e^{\pm \i px}$ must obey $p^2=p_\mu p^\mu=m^2$
\begin{equation}
\i \bar \Gamma^\nu \partial_\nu (\i \Gamma^\mu \partial_\mu \Psi ) = -\Box \Psi = m^2 \Psi = 
\i \bar \Gamma^\nu \partial_\nu (mS\Psi) 
\end{equation}
the transformed bivector fulfills the wave equation
\begin{equation}
\i \bar \Gamma^\mu \partial_\mu (S\Psi) - m \Psi=0. \label{trafoeq}
\end{equation}
Acting on equation (\ref{trafoeq}) with $S$ leads to the requirement
\begin{equation}
S \i \bar \Gamma^\mu \partial_\mu (S\Psi) =  m S \Psi = \i \Gamma^\mu \partial_\mu \Psi
\end{equation}
or
\begin{equation}
S \i \bar \Gamma^\mu S = \i \Gamma^\mu \, . \label{cond}
\end{equation}

If $S$ is a linear operator, it must fulfill
\begin{equation}
S^2= {\Id}_3 \, , \quad S \tilde \Sigma_k S =- \tilde \Sigma_k \, , \quad k=1,2,3 
\end{equation}
or
\begin{equation}
S \tilde \Sigma_k S^{-1} = -\tilde \Sigma_k \, , \quad k=1,2,3 \, .
\end{equation}
This is impossible, since the structure constants of Lie algebra a stable under similarity
transformations.

\noindent If $S$ is anti-linear, it can be written as $S=\tilde S K$, where K denotes complex conjugation.
The complex conjugation of the imaginary unit in equation (\ref{cond}) et cetera then leads to
\begin{equation}
\tilde S \tilde S^*=- {\Id}_3 \, , \quad \tilde S \tilde \Sigma^*_k \tilde S^* = \tilde \Sigma_k \, , \quad k=1,2,3 
\end{equation}
or
\begin{equation}
\tilde S \tilde \Sigma^*_k \tilde S^{-1} = -\tilde \Sigma_k \, , \quad k=1,2,3 \, .
\end{equation}
Again, no such $\tilde S$ exists in three dimensions, since $det ( S \tilde S^*) = det(\tilde S)
det(\tilde S^*)=det(\tilde S) det(\tilde S)^*>0$ contradicts $det({-\Id}_3)=-1$.
This can be contrasted with the case in two dimensions, where
the Pauli matrices $\vec{\sigma}=(\sigma_1, \sigma_2, \sigma_3)$ obey
\begin{equation}
\epsilon \vec{\sigma}^* \epsilon^{-1} = -\vec{\sigma}
\end{equation}
where $\epsilon$ is the totally antisymmetric tensor in two dimensions
\begin{equation}
\epsilon=\left(\begin{array}{cc}
0 & 1 \\
-1 & 0 
\end{array}\right)\; \, , \quad \epsilon^2= (\eta \epsilon)(\eta \epsilon)^*=-\Id_2 \, , \quad det(\epsilon)=det(-\Id_2) 
\end{equation}
such that the Majorana equation(s) as  a generalization of the Weyl equations \cite{Weyl}
and Lorentz invariant two-component field equations
describing a massive particle exist \cite{Majorana,Aste}
\begin{equation}
\i \sigma^\mu \partial_\mu \Psi - m \eta \epsilon \Psi^*=0 
\end{equation}
where $\sigma^\mu = ({{\Id}}_2, \vec{\sigma})$ and $\eta$ is a phase.

\subsection{Mass Term II: "Gauge Formalism"}
One might try to invoke a mass term by naively introducing an ${\rm{SO}}(3,\mathbb{C})$ bivector "gauge potential"
\begin{equation}
H=H^R+ \i H^I=
\left(\begin{array}{r}
H^R_1+ \i H^I_1 \\
H^R_2+ \i H^I_2\\
H^R_3+ \i H^I_3
\end{array}\right) \; 
\end{equation}
related to the massive bivector field via
\begin{equation}
\Psi= \i \bar \Gamma^\nu \partial_\nu H=( \i \partial_0 + \nabla \times) (H^R + \i H^I)
=(-\dot{H}^I + \nabla \times H^R)+ \i(\dot{H}^R+ \nabla \times H^I) 
\end{equation}
fulfilling the massive wave equation
\begin{equation}
\Box H + m^2 H =0.
\end{equation}
This would imply
\begin{equation}
\i  \Gamma^\mu \partial_\mu ( \i \bar \Gamma^\nu \partial_\nu H)=
(- \partial_0^2-\nabla \times \nabla \times) H=
(- \partial_0^2  +\Delta  -\nabla  (\nabla \cdot)H= -m^2 H
\end{equation}
and therefore $\nabla(\nabla \cdot H)=0.$ This condition is, however, not Lorentz invariant.\\

\noindent One has to accept that a mass term is impossible for simple group theoretical reasons. Since a four-gradient
transforms according to the $(\frac{1}{2},\frac{1}{2})$-representation of the Lorentz group, it produces
quantities transforming according to the representations $(\frac{1}{2},\frac{1}{2}) \otimes (1,0)=(\frac{3}{2},\frac{1}{2}) \oplus
(\frac{1}{2},\frac{1}{2})$ when acting on a bivector under $(1,0)$. In the case of Majorana fermions,
one has $(\frac{1}{2},\frac{1}{2}) \otimes (\frac{1}{2},0)=(1,\frac{1}{2}) \oplus
(0,\frac{1}{2})$, such that a the derivative of a field field transforming according to $(\frac{1}{2},0)$ can be coupled
to its complex conjugate field tranforming according to $(0,\frac{1}{2})$.

\section{Coupling to a Current: Transformation Properties}
Adding the real electric current to Amp\`ere's circuital law
\begin{equation}
\frac{\partial E}{\partial t} = +\nabla \times B - {\bf{j}} 
\end{equation}
where a possible factor $\frac{1}{\sqrt{2}}$ according to the normalization chosen in equation (\ref{normdef})
and the coupling constant have been absorbed in ${\bf{j}}$,
shows that equation (\ref{nice}) cannot be interpreted as
the "Dirac form" of Maxwell's equations, since the current
${\bf{j}}=(j^1,j^2,j^3)$ consists of three spatial components of a charge-current four-vector density
\begin{equation}
j=\{j^{\mu}\}=(j^0,j^1,j^2,j^3)=(\rho,j^1,j^2,j^3) \, .
\end{equation} 
Equation (\ref{fun}) becomes
\begin{equation}
\frac{\partial \Psi}{\partial t}  = -\i \cdot  \nabla \times \Psi - {\bf{j}}  \label{fun2}
\end{equation}\\
\noindent and taking the divergence of equation (\ref{fun2}) leads to
\begin{equation}
\nabla \cdot \dot{\Psi} = -\i \cdot \nabla \cdot (\nabla \times \Psi) - \mbox{div} \, {\bf{j}}=
 \dot{\rho}  \label{cont}
\end{equation}
since the continuity equation $\dot{\rho}+ \mbox{div} \,  {\bf{j}} = 0$ holds. Due to the
absence of magnetic charges, equation (\ref{cont}) is equivalent to $div \, \dot{E} = \dot{\rho}$.\\

\noindent Therefore, although the bivector field only couples to the spatial components ${\bf{j}}$
of the charge-current four-vector density $(\rho,{\bf{j}})$, the
charge distribution is encoded in the divergence of the bivector itself
and does not appear as an independent dynamical variable, since the current density $j$ together with the
initial conditions for the charge distribution fix the actual charge density or the divergence of
the bispinor at any time.\\

\noindent Considering the transformation properties of a Dirac spinor under a Lorentz transformation $x'=\Lambda x$
for a moment
\begin{equation}
\psi'(x')=Q_D(\Lambda) \psi(x) \, , \quad \Lambda \in {\rm{SO}}^+(1,3) \, , \quad x'^\mu=
\Lambda^\mu_{\, \, \nu}  x^\nu
\end{equation}
we observe that the Dirac equation
\begin{equation}
\i \gamma^{\mu} \partial_\mu \psi - m \psi = 0 
\end{equation}
also holds in the primed coordinate system. Namely, requiring
\begin{equation}
\i \gamma^\mu \partial'_\mu \psi'(x') =  \i \gamma^\mu \partial'_\mu Q_D \psi(x) = m Q_D \psi(x)
\end{equation}
implies
\begin{equation}
\i \gamma^\mu \Lambda_\mu^{\, \, \nu} \partial_\nu Q_D \psi(x) = m Q_D \psi(x)
\end{equation}
or
\begin{equation}
\Lambda^\alpha_{\, \, \nu}  \Lambda_\mu^{\, \, \nu} Q_D^{-1} \gamma^\mu Q_D =
\Lambda^\alpha_{\,\, \nu} \gamma^\nu \, . \label{prestep}
\end{equation}
Using $\Lambda^\alpha_{\, \, \nu}  \Lambda_\mu^{\, \, \nu}= \delta^{\alpha}_{ \mu}$, we obtain
\begin{equation}
Q_D^{-1} \gamma^\alpha Q_D = \Lambda^\alpha_{\, \, \nu} \gamma^\nu \, . \label{manifest}
\end{equation}
This important property of the Dirac matrices which expresses the manifest
Lorentz covariance of the Dirac equation
holds if the transformation of the spinor components is performed
by an appropriately chosen matrix $Q_D$.\\

\noindent However, a simple relation analogous to equation (\ref{manifest}) \`a la
\begin{equation}
Q^{-1} \Gamma^\alpha Q = \Lambda^\alpha_{\, \, \nu} \Gamma^\nu 
\end{equation}
or
\begin{equation}
Q^{T} \Gamma^\alpha Q = \Lambda^\alpha_{\, \, \nu} \Gamma^\nu \, ,  \quad
Q^{*} \Gamma^\alpha Q = \Lambda^\alpha_{\, \, \nu} \Gamma^\nu
\end{equation}
does not hold, as can be shown by a straightforward calculation.

\noindent In the bivector case, we have
\begin{equation}
\i \Gamma^\mu \partial_\mu \Psi = -\i {\bf{j}}
\end{equation}
and therefore in a primed coordinate system
\begin{equation}
\i \Gamma^\mu_{ab} \partial'_\mu \Psi'_b(x') = - \i {\bf{j}}'_a(x') \, 
\end{equation}
implying (with latin indices $a$, $b$ \ldots = 1,2,3 denoting bivector indices or components
of the current density)
\begin{equation}
\i  \Gamma^\mu_{ab} \Lambda_\mu^{\, \, \nu} \partial_\nu Q_{bc} \Psi_c(x) = -\i \Lambda^{a}_{\, \, \nu}
j^\nu(x) =  -\i \Lambda^{a}_{\, \, 0} \,  div \, E(x) - \i \Lambda^{a}_{\, \, c} \, {\bf{j}}^c (x)
\label{comp}
\end{equation}
since $ j^\mu(x)=(div \, E, j^1, j^2, j^3) \, .$\\

\noindent The divergence of the electric field can be written by the help of the Kronecker delta as
\begin{equation}
div \, E = \delta^{\nu}_{ c} \partial_\nu \Psi_c 
\end{equation}
since the magnetic field is divergence free from the start.
Note that the absence of magnetic charges is not relevant for the present considerations.
Hence equation (\ref{comp}) becomes
\begin{equation}
\i ( \Gamma^\mu_{ab} \Lambda_\mu^{\, \, \nu} Q_{bc} + \Lambda^a_{\, \, 0} \delta^{\nu}_{ c}  )
\partial_\nu \Psi_c(x) = - \i \Lambda^a_{\, \, c} {\bf{j}}^c  \label{final}
\end{equation}
Defining $\tilde{\Lambda}$ as the inverse of the 3-by-3 submatrix $\Lambda^a_{\, \, b}$ according to
$\tilde{\Lambda}^{d}_{\, \, a} \Lambda^a_{\, \, b}= \delta^{d}_{b}$, one may multiply equation (\ref{final})
by $\tilde{\Lambda}^d_{\, \, a}$ and obtains the original equation, i.e.
\begin{equation}
\i (\tilde{\Lambda}^d_{\, \, a} \Gamma^\mu_{ab} \Lambda_\mu^{\, \, \nu} Q_{bc} +
\tilde{\Lambda}^d_{\, \, a}  \Lambda^a_{\, \, 0} \delta^{\nu}_{ c})
\partial_\nu \Psi_c(x) = -\i \tilde{\Lambda}^d_{\, \, a} \Lambda^a_{\, \, c} {\bf{j}}^c = - \i {\bf{j}}^d 
\end{equation}
and therefore
\begin{equation}
\Gamma^\nu_{dc} = \tilde{\Lambda}^d_{\, \, a} \Gamma^\mu_{ab} \Lambda_\mu^{\, \, \nu} Q_{bc} +
\tilde{\Lambda}^d_{\, \, a}  \Lambda^a_{\, \, 0} \delta^{\nu}_{ c} \, . \label{bispi}
\end{equation}
Restricting our consideration to spatial rotations, we have
$\Lambda^a_{\, \, b} = Q_{ab}$ and $\tilde{\Lambda}^{-1} = Q^{-1}$, furthermore
$\Lambda^{a}_{\, \, 0}=0$ for $a=1,2,3$. Accordingly,
\begin{equation}
\Gamma^\nu =  \Lambda_\mu^{\, \, \nu} Q^{-1} \Gamma^\mu Q 
\end{equation}
in analogy to equation (\ref{prestep}),
since for pure rotations, spatial derivatives, the bivector and  the current density transform as vectors
under ${\rm{SO}}(3)$, and the complicated situation in equation (\ref{bispi}) arising from coupling a bivector
to a four-vector does not arise.

\section{Conclusions}
In this paper, the main features of  the gauge free formalism describing a massless spin-1 field coupled to
a conserved current based on the representation of the proper (also called \emph{restricted}) Lorentz group
${\rm{SO}}^+(1,3)$ by the complex orthogonal group ${\rm{SO}}(3,\mathbb{C})$ are investigated. It is shown
that a mass term analogous to the Majorana or Dirac case is impossible for a bivector field for
group theoretical reasons, although
the equations of motion in the bivector formalism display some commonalities with the spinor formalism. It is hoped
that the paper fills a gap in the literature concerning the representation theory of the Lorentz group
in low dimensions relevant for relativistic field theory. Still, the present formalism based on electric and
magnetic fields is unable to describe the coupling to charged fields in a manifest local manner and therefore is
of limited practical use.


\begin{thebibliography}{99}\itemsep=-.2pc

\bibitem{Aste} Aste A., 
{\it A Direct Road to Majorana Fields},
Symmetry {\bf{2}} (2010) 1776--1809.


\bibitem{Belinfante} Belinfante F. and Swihart J.,
{\it Phenomenological Linear Theory of Gravitation: Part III: Interaction with the Spinning Electron},
Annals Phys. {\bf{2}} (1957) 81--99.


\bibitem{Birula} Bialynicki-Birula I.,
{\it Photon Wave Function},
Prog. Opt. {\bf{36}} (1996) 245--294.


\bibitem{Dirac} Dirac P.,
{\it The Quantum Theory of the Electron},
Proc. Roy. Soc. {\bf{A117}} (1928) 610--624.


\bibitem{Majorana} Majorana E.,
{\it Teoria Simmetrica dell'Elettrone e del Positrone
(Symmetric theory of the electron and the positron)},
Nuovo Cim. {\bf{14}} (1937) 171--184.


\bibitem{Silberstein} Silberstein L.,
{\it Elektromagnetische Grundgleichungen in Bivektorieller Behandlung},
Ann. Phys. (Leipzig) {\bf{327}} (1907) 579--586.


\bibitem{Sommerfeld}
Sommerfeld A., {\it Vorlesungen \"uber Theoretische Physik. Band 3: Elektrodynamik},
Dieterichsche Verlagsbuchhandlung, Wiesbaden 1948.


\bibitem{Weber}
Weber H.,
{\it Die Partiellen Differential-Gleichungen der Mathematischen Physik nach Riemann's Vorlesungen},
Vol. 2, Friedrich Vieweg und Sohn, Braunschweig 1901.


\bibitem{Weyl} Weyl H.,
{\it Elektron und Gravitation},
Z. Phys. {\bf{56}} (1929) 330--352; $\,$
{\it Gravitation and the Electron},
Acad. Sci. U.S.A. {\bf{15}} (1929) 323--334.

\bibitem{Wigner} Wigner E.,
{\it On the Unitary Representations of the Inhomogeneous Lorentz Group},
Ann. Math. {\bf{40}} (1939) 149--204. 

\end{thebibliography}
\end{document}